\documentclass{aa}
\usepackage{graphicx}
\usepackage{natbib}
\begin{document}
\title{Investigating the
central engine of Seyfert 2 galaxies \textit{with} and \textit{without} Polarized Broad Lines}
\author{S. J. Deluit
\inst{1,2}}
\institute{\textit{INTEGRAL} Science Data Center, 16 Chemin d'Ecogia, ch-1290 Versoix, Switzerland
\and Geneva Observatory, 51 Chemin des Maillettes, ch-1290 Sauverny, Switzerland}
\offprints{S. J.  Deluit,\\ \email{Sandrine.Deluit@obs.unige.ch}}
\date{Received 28 April 2003/Accepted 9 October 2003}
\abstract{We study  the  hard X-ray  emission of two samples of Seyfert 2 galaxies \textit{with} and \textit{without} Polarized Broad Lines (PBL). In the hard X-ray domain, absorption effects do not significantly modify  the intrinsic emission
 allowing
 us a direct
 access
 to the central engine.\\
The purpose of this study is  to compare the primary  emission of the two Seyfert 2  subclasses in order to investigate
 the nature of their central engine   and to test  unified models according to
 which they both have 
a hidden Seyfert 1 nucleus.  We compute the average hard X-ray spectra of Seyfert 2 galaxies with and without PBL
 observed
 with BeppoSAX/PDS (15-136 keV).
 The two spectra have a  common general behavior at  first sight, but investigating deeper we find  differences in
  the  intrinsic  properties  of  the two categories of Seyfert 2 galaxies. Sy 2
  with polarized broad lines  have
 physical conditions close to those of Sy 1 galaxies whereas Sy 2 without PBL differ substantially, suggesting that
  they may  have a particular  place in the  scheme of Seyfert galaxies. 

}
\titlerunning{The central engine  of Seyfert 2 with and without Polarized Broad Lines}
\authorrunning{S. J. Deluit} 
\maketitle
\section{Introduction}
Seyfert galaxies are classified into two spectroscopic groups based on the presence (type 1)
 or absence (type 2) of 
broad permitted optical emission lines. The first discovery of Polarized Broad Lines (PBL hereafter) in NGC 1068 \citep{Antonucci}
 and later in  several other
 Seyfert 2 galaxies
 \citep{MillerGood,Tran95,Young,Heisler,Awaki,Moran2000,Alexander,GuPbl,Lumsden} suggests
  that  Seyfert 2  
 galaxies harbor a bright Seyfert 1 nucleus hidden from our view by an optically and geometrically thick obscuring
  torus (``Unified models'', see the review of \citet{Antonucci93}). In this frame, the difference between type 1 and 2
   Seyfert galaxies is only due to the viewing
    angle. 
X-ray  observations  support this model showing the presence of large column densities along the line of sight of Seyfert
 2 galaxies \citep {Risaliti, Guainazzi2001}.
However, other observations are inconsistent with   unified models, in particular in the hard X-ray domain where \citet{Z95} and
 \citet{Deluit1} (D03 hereafter)  show differences in the spectral index, energy cutoff and amount of
  reflection between
  Sy 1 and Sy 2 galaxies. Furthermore, \citet{Alexander} refutes the absorption argument to explain the absence
   of broad lines in some Sy 2 galaxies  showing that the density of the absorbing medium is not linked to the detectability of broad lines.\\
   The polarization is presumed to  be due to nuclear emission scattered by  a gas of electrons located further out
    than the obscuring material. According to current  models, the detection of polarized broad emission lines in type
     2 Seyfert  galaxies
  implies that these objects have a Hidden Broad Line Region (HBLR henceforth).
 However, not all Sy 2 galaxies that have been observed in polarized light show the presence of  broad lines.
  This might imply an 
intrinsic difference
 between Seyfert 2 with PBL (PBLs hereafter) and those with  No PBL (NPBLs hereafter). We propose to
  test this by studying their hard X-ray intrinsic emission in a spectral domain where absorption does not modify
   significantly the primary emission.\\
The existence of Sy 2 without PBL could be due to  the absence of a broad line region  in some Sy 2 galaxies or
 to the absence of a scattering medium or different physical conditions within the inner regions.  \\
To investigate these hypotheses, we intend to describe the general emission properties 
 of various  classes of Seyfert 2
galaxies by computing their average hard X-ray spectrum.\\
In D03 we computed the average spectra of Sy 1, Sy 1.5 and Sy 2 galaxies. We found that  the Sy 1 and Sy 2
 galaxies  of our sample have
  a different behavior,  in
particular concerning  the presence of a cutoff which is required in the Sy 1 emission and absent in the Sy 2
average spectrum. Another point of disagreement was the role played by  Compton reflection
possibly influencing  the Sy 2 emission, unlike Sy 1. \\
We  pursue our investigations  within the Seyfert
 2 class of our sample to search for common properties between Sy 1 galaxies and the two Sy 2 subclasses. To test the
  argument that the absorption is the only reason for which we cannot see the HBLR, we  
 consider Compton-thin Sy 2 galaxies. 
  After presenting  the
  average spectra obtained, we study physical processes  
responsible for the observed X$\gamma$ emission.
\section{The Dataset}
The initial sample of Seyfert 2 galaxies is that  presented in D03.
 From the initial 22 Seyfert 2 galaxies we keep 16 objects for a total of 24 observations.
 Five objects have been observed several times.
We apply selection criteria, presented in subsections 2.1. and 2.2., to the initial sample.
\subsection{Spectropolarimetric data}
 This study is based on a sample composed of Seyfert 2 galaxies of two types: one presenting broad
 lines in polarized light, the second for which no PBL have been detected. We point out the fact that the
  non detection of broad lines does not prove definitively their genuine absence, in particular if we consider
   the limitations of  current spectropolarimetric instrumentation. However, the purpose of this study is
    to  start  a  comparison  of the two subclasses of Seyfert 2 galaxies in the
    hard  X-ray  domain,  using  the  best available published information on the
    presence of hidden broad lines. \\
The  current  classification of Seyfert galaxies is mainly determined by optical
criteria, summarized in Table 1. \\
We collected all Seyfert 2 galaxies from the recent literature (from 1980 to 2003)
for which   
 spectropolarimetric data are available. 
Several objects of the D03 Sy 2 sample have been excluded: IRAS 18325-5926  because of  its NED classification as 
Sy 2 which has been questioned   by
\citet{Iwasawa}; NGC 4939 is  a Sy 2 but no spectropolarimetric data is available.
 NGC 5674 and  ESO 103-G35 have been excluded because of their Sy 1.9 classification.  Indeed, current models 
  generally  claim  that  the  detection  of  a broad H$\alpha$ line implies the
  presence of a hidden broad line region and
  explain
    the  absence of the  H$\beta$ line  as an obscuration of this region, leading to the conclusion that Sy 1.9 galaxies
     harbor  a hidden Sy 1 nucleus. \\
In addition to spectropolarimetric considerations, we use only data with  an integrated signal to noise ratio higher
 than 2  in
 the PDS energy range (15-136 keV). This threshold is high enough to conclude  a detection by the PDS detector for a
  known source. 
We therefore  consider in our study all Sy 2 galaxies for which both spectropolarimetric information and   PDS data
of sufficient  quality are available.
\begin{table}[width=4cm]
\caption{Classification of Seyfert galaxies by  optical criteria}
\begin{flushleft}
\begin{tabular}[width=4cm]{l|l} 
%\begin{table}{l l}
\hline
\hline
Classification & Optical Properties\\
\hline
Seyfert 1 galaxies& broad Balmer lines\\
Seyfert 1.5 galaxies & Apparent narrow H$\beta$ profile \\ 
&  superposed on broad wings\\
Seyfert 1.9 galaxies & Broad component visible in H$\alpha$ \\
&  but not in  H$\beta$ \\
Seyfert 2 with PBL  & Broad component visible only in \\
  &  polarized Balmer lines \\
Seyfert 2 without PBL & Broad component invisible with  \\
&any method \\
\hline
%\end{table}
\end{tabular}
\end{flushleft}
\end{table}

\begin{table*}[!hbt]
\begin{center}
\caption{General Characteristics and Spectral Properties of  objects composing the sample}
\begin{tabular}[!hbt]{c c c c l l c c c}
\hline
\hline
\large{Source name} &
\large{RA}   &
\large{DEC}  &
\large{Redshift}	&
\large{~~~N$_{\footnotesize{\textrm{H}}}$}$^{1}$ &   
\large{~~~PBL } &
\large{S/N}$^{4}$ &
\large{F$_{15-136~\textrm{keV}}$}$^{5}$ &
\large{L$_{15-136~\textrm{keV}}$}$^{6}$ \\
  &(\it{h  m  s})& ($^{\circ}$  $^{\prime}$  $^{\prime\prime}$)   & z & & ~~presence?& & & \\

\hline
IRAS 00198-7926 & 00 21 53.8 & $-$79 10 08.0 & 0.073 & ~~~~~ - & ~~~~N\tiny{$^{[L]}$} & 2.10 & 2.18 &
22.9 \\
NGC 1358 & 03 33 39.5 & $-$05 05 20.0  & 0.013 &~~~~~ - & ~~~~N\tiny{$^{[Mb]}$} & 2.28 & 1.38 & 0.48 \\
IRAS 05189-2524 & 05 21 01.3 & $-$25 21 42.9  & 0.043& ~~4.90\tiny{$^{[B]}$}  &  ~~~~Y\tiny{$^{[Y]}$} & 2.66 & 1.14 & 4.04\\
NGC 2110 & 05 52 11.2 & $-$07 27 20.8  & 0.008 & ~~2.89\tiny{$^{[H]}$}  &   ~~~~Y\tiny{$^{[VC]}$}  & 20.5 & 8.62 & 1.01 \\				  
NGC 2992$^{3}$& 09 45 42.0 & $-$14 19 36.9 & 0.008& ~~0.69\tiny{$^{[W96]}$}  & ~~~~N\tiny{$^{[K, VC]}$} &
41.0 & 11.0 & 1.35 \\
MCG 5-23-16 & 09 47 40.1 & $-$30 56 53.9 & 0.008 & ~~1.62\tiny{$^{[W97]}$} & ~~~~Y\tiny{$^{[Go]}$} & 45.8 & 19.8 & 2.62 \\
IRAS F12072-0444 & 12 09 45.1 & $-$05 01 14.9  & 0.128 &~~0.17\tiny{$^{[D]}$} &  ~~~~N\tiny{$^{[V]}$}  & 4.88 & 1.84 & 61.3  \\
NGC 4388$^{2}$ & 12 25 46.7 & +12 39 44.0 & 0.008& ~~42.0\tiny{$^{[B]}$} & ~~~~Y\tiny{$^{[Y]}$} & 63.8 & 22.1 &
3.00\\
NGC 4507$^{3}$ & 12 35 37.0 & $-$39 54 32.0 & 0.012 &~~29.2\tiny{$^{[Co]}$} &
~~~~Y\tiny{$^{[Ma]}$} & 58.1 &
19.1 & 5.12 \\
NGC  5252  &  13  38  16.3  &  +04  32  20.0  &  0.023 &~~4.33\tiny{$^{[Ca]}$} &
~~~~N\tiny{$^{[VC]}$} & 2.86 & 2.26 & 2.34\\
NGC 5506$^{3}$ & 14 13 14.8 & $-$03 12 28.0 & 0.006  & ~~3.40\tiny{$^{[S]}$}   &
~~~~Y\tiny{$^{[VC]}$}  & 69.0 & 17.0 & 1.25 \\
NGC 6300 & 17 17 00.3 & $-$62 49 15.0 & 0.004 & ~~60.0\tiny{$^{[Le]}$}         & ~~~~N\tiny{$^{[Le]}$} &23.1 & 8.35 & 0.22  \\
IRAS 20210+1121 & 20 23 25.6 & +11 31 32.9 & 0.056 & ~~$\leq$6.0\tiny{$^{[U]}$} & ~~~~Y\tiny{$^{[Y]}$} & 2.24 & 0.87 & 5.45 \\
NGC 7172$^{2}$ & 22 02 02.2 & $-$31 52 12.0 & 0.009 &~~8.61\tiny{$^{[G]}$} &
~~~~N\tiny{$^{[H]}$} & 9.73 & 4.43
& 0.65 \\
IRAS 23060+0505 & 23 08 33.8 & +05 21 29.0 & 0.173 &~~8.40\tiny{$^{[Br]}$} &   ~~~~Y\tiny{$^{[Y]}$}  & 4.20 & 1.79 & 111 \\
NGC 7582 & 23 18 23.5 & $-$42 22 11.9  & 0.005 & ~~12.4\tiny{$^{[X]}$} &   ~~~~N\tiny{$^{[H]}$} & 24.2 & 10.5 & 0.57\\
\hline
\end{tabular}
\end{center}
Notes:\\
$^{1}$ 10$^{22}$ cm$^{-2}$, 
$^{2}$ 2 observations, 
$^{3}$ 3 observations,  
$^{4}$ Integrated signal to noise ratio in the PDS energy range, 
$^{5}$ Integrated flux in 10$^{-11}$ erg~s$^{-1}$~cm$^{-2}$, 
$^{6}$ Integrated luminosity in 10$^{43}$ erg~s$^{-1}$\\
  \textit{References concerning the Hydrogen column density:} \\
	B: \citet{Bassani}, 
	Br: \citet{Brandt},  
	Ca: \citet{Cappi}, 
	Co: \citet{Comastri}, 
	D: Deluit et al., Paper II of the D03 study, 
	G: \citet{Guainazzi98},  
	H: \citet{Hayashi}, 
	Le: \citet{Leighly}, 
	S: \citet{SmithDone} , 
	U: \citet{Ueno}, 
	W96: \citet{Weaver96},  
	W97: \citet{Weaver97}, 
	X: \citet{Xue} \\
\textit{References concerning the spectropolarimetric  information:} \\
Go: \citet{Goodrich},
H: \citet{Heisler}, 
K: \citet{Kay},  
L: \citet{Lumsden}, 
Ma: \citet{Moran2000}, 
Mb: \citet{Moran2001}, 
V: \citet{Veilleux}, 
VC: \citet{Veron}, 
Y: \citet{Young}

\end{table*}     

 \subsection{Sources not significantly influenced by absorption effects}
As exhaustively  explained in D03, we limit our study to  Sy 2 galaxies having a hydrogen column density less
 than 7$\cdot$10$^{23}$cm$^{-2}$. We have established that below this threshold, the absorption effects do not 
 alter strongly the primary emission, allowing us  a direct view of the  emission produced by the
   central engine.
\subsection{Sample spectral characteristics}
 The final sample is composed of 8 Seyfert 2 galaxies with PBL and 8 without PBL detected (Table 2).\\

\subsubsection{Absorption distribution}
We present in Figure 1 the hydrogen column density distribution of PBLs  and NPBLs  
Sy 2 galaxies.
\begin{figure}[!h]
\centering
\includegraphics[width=8cm]{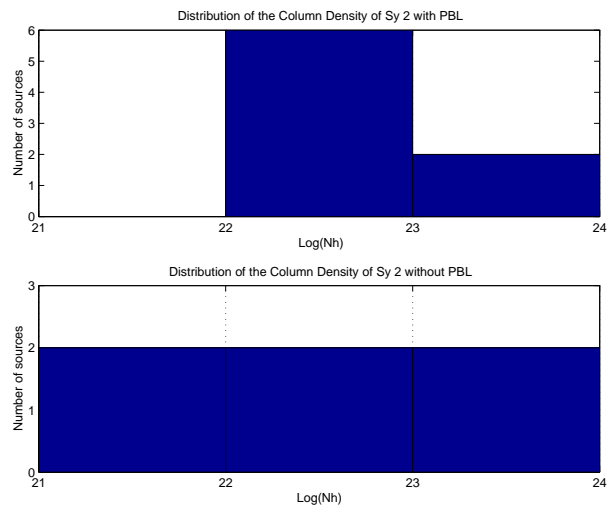}
\caption{Hydrogen column density distributions of Sy 2 with polarized broad lines (upper panel) and without PBL
 (lower panel)}
\end{figure}
Several contradictory results have been found during  recent years. \citet{Heisler} and \citet{GuPbl}  found a
higher column density for NPBLs whereas \citet{Alexander} found no link between the detectability of polarized
 broad lines and the density of the absorbing material. \\
In our study,  the column density distributions  of the two Sy 2 classes are not significantly different. In particular
 the NPBLs galaxies are not  more absorbed  than PBLs since they have the same number of objects with a low absorption
  (N$_{H}$$<$10$^{22}$cm$^{-2}$) than with a higher absorption (N$_{H}$$>$10$^{22}$cm$^{-2}$). 
  The fact that we consider  a small number of objects
  and only Compton-thin Sy 2 galaxies prevents us from generalizing this result for the overall Sy 2 class.  We
   however show  that even in Compton-thin Sy 2 galaxies the absence of polarized broad lines
    is not unusual. 
\subsubsection{Luminosity distribution}
The hard X-ray luminosity is a signature of the nucleus activity. \\
The comparison of luminosity distributions of the two Sy 2 subclasses (Figure 2) reveals that NPBLs are weaker than PBLs, indicating that the activity of their respective central engine is slightly different. Indeed, 87.5 \% of PBLs have a luminosity between 10$^{43}$ and  10$^{44}$ erg~s$^{-1}$ against only 25\% of NPBLs. 50\% of the NPBLs population has a luminosity between 10$^{42}$ and  10$^{43}$ erg~s$^{-1}$. \\
 The hypothesis according to which  PBLs would be more luminous than NPBLs has been evoked in other wavelength domains (e.g. \citet{Kay} and \citet{Tran2003}); we discuss the implications of  this result in the section 7.

\begin{figure}[!h]
\centering
\includegraphics[width=8cm]{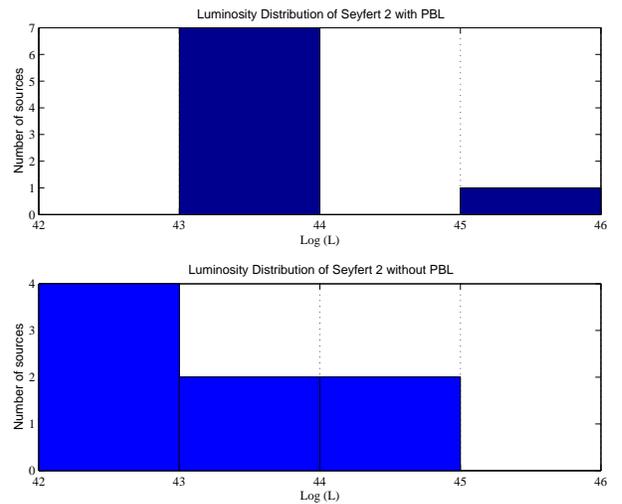}
\caption{Luminosity  distributions of Sy 2 with polarized broad lines (upper panel) and without
PBL (lower panel)}
\end{figure}
\section{Data Analysis}
The Narrow Field Interments (NFI) of the BeppoSAX satellite  is composed of a Low Energy Concentrator Spectrometer
 (LECS; \citep{Parmar}), three
Medium Energy Concentrator Spectrometers (MECS; \citet{Boella}) and a Phoswich Detector
Counter (PDS; \citet{Frontera}). LECS and MECS have imaging capabilities and operate
respectively in the 0.1-5 keV and 2.0-10 keV domains. The PDS, used in our study,  is an instrument functioning in an
energy band between 15-200 keV with an energy resolution of 15\% at 60 keV. 
The detector is designed to  allow a good control of background
variations using rocking collimators that periodically sample source+background combinations and
background alone.\\
The data analysis has been performed with the  XSPEC \citep{Arnaud}  version 11  package  using the latest PDS 
response matrices
released by the BeppoSAX SDC\footnote{http://www.asdc.asi.it/bepposax/}. 
To  compute  the  average spectra, we used the task Mathpha version 5.1.
combining the data files. The errors
 have  been  propagated at each step of the procedure. In this procedure, we did
 not consider
the background since the individual spectra were  already background subtracted. 
 We computed the average count spectrum weighting
 the individual spectra by their respective exposure time. \\
We kept the same binning for the average spectrum as  for the  original data of each source. 
For a source observed several times, we first computed its average  spectrum.  
 The PDS energy band used  in our study extends from 15 to 136 keV.
\section{Spectral properties of various Seyfert 2 galaxies classes}  
We present in Figure 3 the average spectrum of Seyfert  2 galaxies with and without polarized broad lines.

\begin{figure}[!h]
\centering
\includegraphics[width=6.5cm, angle=-90]{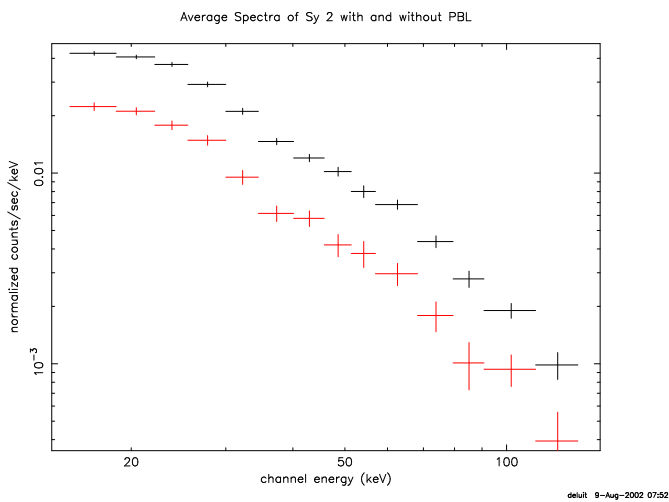}
\caption{Average spectra of Seyfert 2 galaxies with (black) and without  
(light red) polarized broad lines} 
\end{figure}

The integrated signal to noise ratio in the 15-136 keV energy range is 
 98 for the PBLs and  47 for the NPBLs average spectrum.
The NPBLs countrate is half as strong as the PBLs at 15 keV. \\
We    
 compare  the  central engine signatures of the two types of Sy 2 galaxies using
 two methods: firstly, with a model-independent analysis comparing  directly the normalized count
 spectra  independently  of  any  instrumental  consideration;  secondly, with a
 model-dependent method using a spectral
 fitting procedure. \\
 We also intend to compare the emission of the 
   two Sy 2 classes   with the Sy 1 galaxies to investigate whether they have common properties as supposed by unified
     models. The Seyfert 1 average spectrum used in this study
    comes
    from D03.
\subsection{Model-Independent Analysis: comparison of the normalized count spectra }
We compare the average count  spectra normalizing them at 15 keV (Figure 4). 
\begin{figure*}[!hbt]
\centering
\includegraphics[ width=11cm]{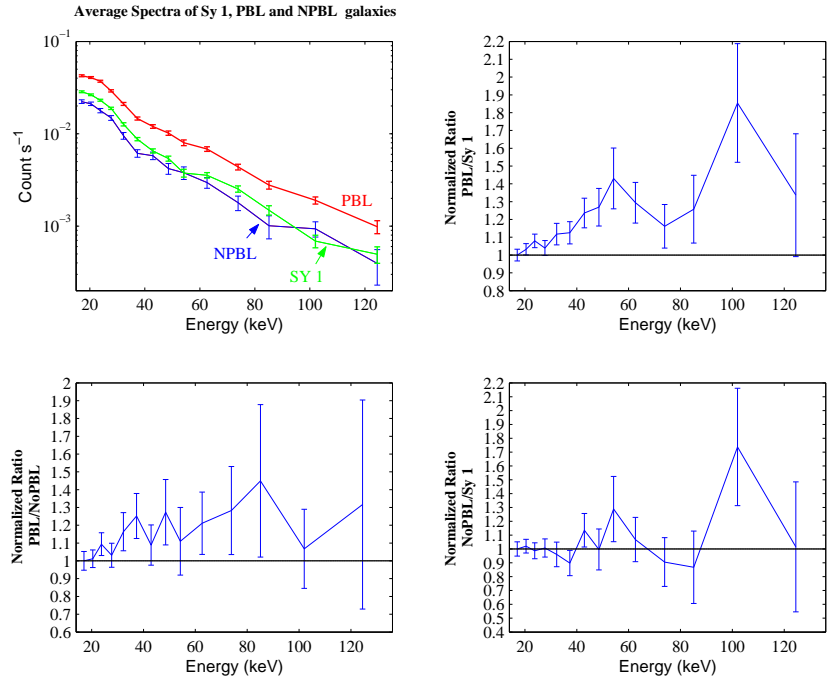}
\caption{Left: average spectra of Sy 1, PBLs and NPBLs  galaxies (upper panel) and the normalized ratio of
 Spectral Energy Distributions of the  pair PBLs/NPBLs (lower panel). Right: ratios of
 SEDs of the pairs: PBLs/Sy 1 and NPBLs/Sy 1 galaxies}
\end{figure*}
%\end{flushleft}
\subsubsection{Comparison between Seyfert 2 galaxies with and without polarized broad lines}
 The average spectra of PBLs and NPBLs could be almost identical considering the error bars. Nevertheless, the PBLs emission seems to be slightly higher than that of NPBLs.  To strengthen this result, we rebin 
the spectra to improve the bin 
  significance (Figure 5).
\begin{figure}[!hbt]
\centering
\includegraphics[width=6cm]{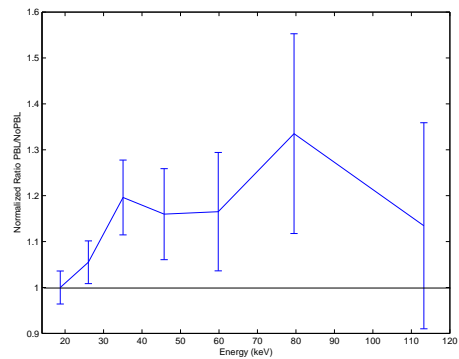}
\caption{Normalized ratio of rebinned spectra  of the pair PBLs/NPBLs}
\end{figure}  
The PBLs emission is higher than NPBLs along the major part of the energy domain indicating that 
the PBLs spectrum  contains 
 a larger hard photons population than NPBLs. 
The PBLs average spectrum is 45\% higher than that of NPBLs   around 85 keV where they differ most in the 
 ratio of Figure 4. 
\subsubsection{Comparison of PBLs and NPBLs Sy 2 galaxies with Sy 1 galaxies  }
 \begin{enumerate}
\item \textit{Comparison of Sy 2 with PBL and Sy 1} \\
The PBLs emission is always higher than that of  Sy 1 galaxies giving an indication of the hardness
  of this category of Sy 2 galaxies.
 We note two  excesses  of  the PBLs emission compared with Sy 1 around 55 keV and  100 keV.  The deficit of hard
  photons in the Sy 1 class can be explained either
   by the presence of a  cutoff in its emission  or by a lower PBLs \textit{intrinsic} spectral index
and/or  by a reflection component in PBLs.  
The PBLs emission is 85\% greater
 than  Sy 1 at $\sim$100 keV.
\item \textit{Comparison of Sy 2 without PBL and Sy 1} \\
The two spectra are similar below 40 keV confirming that the emission of NPBLs is not significantly modified
 by absorption effects at these energies.
The average spectra  remain close up to $\sim$85 keV.  The NPBLs emission is higher than that of Sy 1 at the
 two  same  energies  (55  and 100 keV) where the PBLs emission dominates in the
 ratio PBLs/Sy 1. This
   suggests the presence of a particular spectral component in the Sy 1 average spectrum. \\  The maximal  difference  measured between the two classes is found at $\sim$100 keV where the NPBLs emission is 74\% 
greater than that of Sy 1.
\end{enumerate}
\subsubsection{Preliminary results}
 At  first  sight  (Fig. 3),  the spectra of Seyfert 2 galaxies with and without polarized broad lines  appear very
similar. However, the model-independent analysis we
performed reveals small but significant differences (Fig. 5).
The comparison with  Sy 1 galaxies  indicates  that singularities exist also between Sy 1 galaxies and the 
two subclasses of Sy 2 galaxies. \\
The physical processes that can be responsible for the observed spectra and for the differences found between
 the various
classes of Seyfert galaxies are investigated in the next subsection.
\subsection{Model-Dependent Analysis: investigating the emission processes}
 To further test the above results, we compare the spectra  using a fitting procedure that will allow
 us  to investigate also the components occurring in the emission. We fit the spectra initially with simple models
  then  with the reflection
  model PEXRAV. 
The results are  shown in  Tables 3 and 4.
\begin{table*}[!hbt]
 \begin{center}
   \caption{Best fit parameters for  simple models}
\begin{tabular}[!hbt]{c c c c c c c  } 
\hline
\hline
 Model & Sy 2 & $\Gamma$ & E$_{cutoff}$ & $\chi^{2}$/d.o.f. &   $\chi^{2}_{r}$ & FTEST \\
  & Class& Spectral index& (keV)  &   & reduced & Probability value \\
\hline
Power Law &  PBL   &  1.85  & - &  43.0/12 & 3.58 & - \\
  & NPBL  &  1.99$_{-0.07}^{+0.08}$ & - &  10.3/12 & 0.86 &  -\\

Cutoff Power Law  & PBL & 1.38$_{-0.15}^{+0.14}$ &  83.6$_{-20.4}^{+35.6}$ & 
09.9/11 & 0.90 &  7.9$\cdot 10^{-5}$ \\
 & NPBL &  1.68$_{-0.33}^{+0.20}$ & 118$_{-60.0}^{+\tiny{\textrm{NC}}}$ &
 07.1/11 & 0.65 &   4.8$\cdot 10^{-2}$  \\
\hline
\end{tabular}
\end{center}
Notes:\\
The uncertainties correspond to 90$\%$
confidence level based on a $\Delta$$\chi^{2}$=2.7 criterion \citep{Lampton}\\
NC indicates that no constraint has been found on this limit.
\end{table*}
\begin{table*}[!hbt]
\begin{center}
 \caption{Best fit Parameters for the reflection model PEXRAV}
\begin{tabular}[!hbt]{c c c c c c c}
\hline
\hline
 Sy 2 & $\Gamma$ & Efolded & R & $\chi^{2}$/d.o.f. & $\chi^{2}$ & cos$\theta$   \\
 Class & Spectral index & (keV) & Reflection & & reduced & fixed \\
\hline
 PBL &  1.65$_{-0.38}^{+0.32}$  &  181$_{-113}^{+\tiny{\textrm{NC}} }$   &  0.51$_{-0.51}^{+1.48}$  & 8.6/10 & 0.86 & 0.45 \\
 NPBL &   2.06$_{-0.68}^{+0.10}$  &  1.7$\cdot 10^{4}_ {\tiny{[a]}}$ &  0.90$_{-0.70}^{+1.87}$  &  6.5/10 & 0.65
  & 0.45 \\
\hline
Sy 2 & $\Gamma$ & Efolded & R & $\chi^{2}$/d.o.f. & $\chi^{2}$ &  cos$\theta$  \\
 Class& Spectral index & (keV) & Reflection & & reduced  &  free  \\
\hline
PBL & 1.62$_{-0.35}^{+0.36}$ &  167$_{-99}^{+\tiny{\textrm{NC}} }$   &   0.27$_{-0.27}^{+\tiny{\textrm{NC}} }$ & 8.6/9 & 0.96 &
  0.81    \\
    NPBL &   2.08$_{-0.10}^{+0.11}$ &  9.7$\cdot 10^{5}_ {\tiny{[a]}}$ &   3.07$_{-3.01}^{+\tiny{\textrm{NC}}}$    & 
     6.5/9 & 0.72 & 0.14  \\
   \hline

  \end{tabular}
\end{center}
 Notes:\\
The uncertainties correspond to 90$\%$
confidence level based on a $\Delta$$\chi^{2}$=2.7 criterion \citep{Lampton}.\\
 NC indicates that no constraint has been found on this limit.\\
$[a]$ the value shows the absence of a measured cutoff in the PDS energy range.
\end{table*}
 \subsubsection{Fitting  of the spectra with simple theoretical models}
We fit the spectra successively with a simple power law  and a cutoff power law (PL and
 CPL hereafter).
To know the best model fitting the spectra,  we refer to the reduced $\chi$$^{2}$  value expected to  be the 
nearest possible to 1. \\
A simple power law cannot reproduce the  PBLs spectrum  ($\chi$$^{2}_{r}$=3.58) thus at least one  additional
 component  is required.  The
  occurrence  of  a  cutoff  in the PBLs observed spectrum is the best scheme.
Indeed, the reduced $\chi$$^{2}$ value ($\chi$$^{2}_{r}$=0.90) for a CPL and in particular
     the
very low   FTEST probability value  (P$_{\textrm{\tiny{FTEST}}}$=7.9$\cdot$10$^{-5}$) indicate the high probability
 of having a  cutoff in the PBLs spectrum.\\
 The NPBLs average spectrum is well represented by a simple power law
and a FTEST gives no justification for including a cutoff in the model.  \\
Considering the best fit obtained for the two Sy 2 classes, NPBLs have a  spectral index greater than PBLs, showing the softness of their
 emission ($\Gamma$$_{\textrm{\tiny{NPBLs}}}^{\textrm{\tiny{PL}}}$=1.99$_{-0.07}^{+0.08}$ against
   $\Gamma$$_{\textrm{\tiny{PBLs}}}^{\textrm{\tiny{CPL}}}$=1.38$_{-0.15}^{+0.14}$    with  a  cutoff energy
    of E$_{c}$=83.6$_{-20.4}^{+35.6}$ keV). \\
 PBLs have the same
 general
  behavior as Sy 1 galaxies,  which also require a cutoff in their average spectrum.
In particular the Sy 1 energy cutoff (E$_{c}$=70$_{-17}^{+30}$ keV, see D03 for spectral parameters) is very similar
 to that of PBLs
 (E$_{c}$=83.6$_{-20.4}^{+35.6}$ keV). 
 The spectral index is possibly slightly steeper for Sy 1 ($\Gamma$=1.49$_{-0.17}^{+0.17}$) than for PBLs  ($\Gamma$=1.38$_{-0.15}^{+0.14}$) but
  considering
  the errors  they could be identical. \\
Therefore, Sy 1 and Sy 2 galaxies \textit{with} polarized broad lines have spectral  similarities whereas NPBLs  galaxies present
 distinct properties  with the absence of a constrained cutoff and a probable greater spectral index.
\subsubsection{Fitting of the spectra with the reflection model PEXRAV}
To investigate if  Compton reflection plays a role in  the PBLs and NPBLs emission we use the PEXRAV model
 \citep{Magdziarz}.  
This model
calculates the expected X-ray spectrum when a point source of X-rays is incident on optically thick,
mainly neutral material. 
The parameter R  is related to the reflection
component and the model is inclination/reflection  dependent.
The results are presented in Table 4. \\
We first fix the inclination from the normal of the disk  at
cos$\theta$=0.45  to compare the various spectral parameters of the two Sy 2 
subclasses independently of any strong influence of the inclination.   The PBLs average spectrum requires a cutoff  whereas no indication of such a presence is observed in NPBLs as evoked  in 4.2.1. 
The spectral indices of Sy 2 with and without PBL are within the same value domain. 
 The presence of a 
 reflection process
in the NPBLs/PBLs emissions is not proven since the PEXRAV model does not improve significantly the fitting results. 
In particular, the reflection process could be totally absent in the PBLs spectrum (R=0.51$_{-0.51}^{+1.48}$). \\
We now leave the inclination free to  limit the dependence between the different spectral parameters. We find that PBLs  would be
 close to  the normal of the  disk with an angle of cos$\theta$=0.81 or 36$^{\circ}$ whereas NPBLs would be seen with
  an 
 angle of   cos$\theta$=0.14 or  82$^{\circ}$. \\
The PBLs angle corresponds exactly to that  found for Sy 1 galaxies in D03 where we
 obtained cos$\theta$=0.81. That does not prove that PBLs and Sy 1 have genuinely  the same inclination but that their spectra give the
  same parameters when 
  applying the  PEXRAV model. 
\section{Comparison of the two different analyses }
The models and  parameters that best fit PBLs  show that their properties are similar to those of 
Seyfert 1 galaxies (see 4.2.). However, the ratio of
 their average count spectra, shown in 4.1.2, indicates a significant difference  at high energies. To understand
  this apparent
 contradiction  between  the  two  methods,  we select  the best fit model for each
 class and we calculate
the ratio of these models (Figure 6).\\
We see in  figure 6b that the PBLs model is higher than that of NPBLs over most  of the energy
 domain  by the same amount ($\sim$30$\%$)  as seen in 4.1.1 in  the count spectra ratio. 
 The PBLs model is always higher than that of
   Sy 1 galaxies (Fig. 6c), that is explained by the Sy 1 cutoff occurring at a slightly lower energy than PBLs.\\
The divergence between NPBLs and Sy 1 galaxies does not exceed 20$\%$ below $\sim$85 keV whereas at high 
energies NPBLs are more powerful emitters (Fig. 6d).  Indeed, the presence of a cutoff in the Sy 1 spectrum at
 $\sim$70 keV entails a rapid decrease	of the Sy 1 emission 
whereas the NPBLs drop in emission is only due to the spectral index of the
continuum.	\\
The model ratios have thus the same general behavior and amplitude as those of the count spectra, confirming
  the results of the fits to the models.
\begin{figure}[!h]
\centering
\includegraphics[width=9cm]{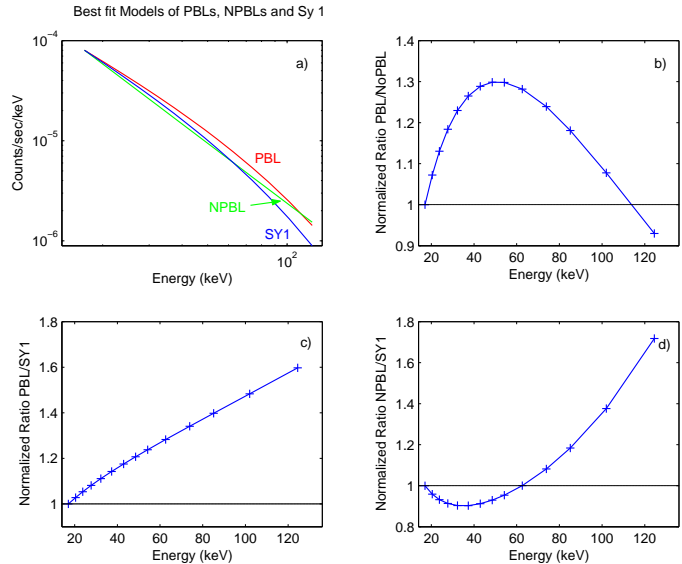}
\caption{Ratios of models corresponding to the best fits. The models are respectively a cutoff power law
 for  PBLs ($\Gamma$$_{\textrm{\tiny{PBL}}}^{\textrm{\tiny{CPL}}}$=1.38, E$_{c}$=83.6 keV) and Sy 1 galaxies ($\Gamma$$_{\textrm{\tiny{Sy1}}}^{\textrm{\tiny{CPL}}}$=1.49, E$_{c}$=70 keV), and a simple power law for NPBLs ($\Gamma$$_{\textrm{\tiny{NPBL}}}^{\textrm{\tiny{PL}}}$=1.99)}
\end{figure}
\section{Results }
 The  spectra  of  our  Seyfert  2 galaxies with and without PBL have a common general
behavior at first sight. Investigating deeper, the PBLs  spectrum presents a well-constrained cutoff  whereas the NPBLs emission does not require this property. \\
  Our results also indicate that Seyfert  2 galaxies with polarized broad lines have several common properties with Sy 1 galaxies (cutoff and
 hardness of the same order).\\
 Therefore, the nature of Seyfert 2 galaxies without polarized broad lines remains unclear. We found significant divergences between NPBLs and the pair  Sy1/PBLs. Indeed,  NPBLs have no constrained cutoff, have a greater spectral index  and are less luminous than PBLs, revealing intrinsic differences. 
\section{Discussion}
Far from having definitively proven that NPBLs have a different central engine hence that they do  not have
 a hidden Sy 1 nucleus, we show
 that their primary emission  presents  substantial differences compared to PBLs and Sy 1 galaxies. 
  \subsection{The high-energy cutoff}
Sy 1 and PBLs galaxies  possess  a cutoff of the same order,   
indicating that the temperature of the electron gas is similar for both classes
  (T$_{\textrm{\tiny{Sy1}}}$=E$_{c}$/k$_{\textrm{\tiny{B}}}$=8.27$\cdot$10$^{8}$ K
 and T$_{\textrm{\tiny{PBLs}}}$=9.88$\cdot$10$^{8}$ K).  
The presence of similar cutoffs in the PBLs and Sy 1 emission reveals that the same global physical conditions 
occur in the inner regions of their source. This supports unified models. \\ 
The absence of a  constrained cutoff in NPBLs emphasizes differences in the 
primary emission of various classes of Sy 2 galaxies. However, a cutoff could arise at higher energies thus produced by very high electron gas temperatures. We therefore 
 have to extend the energy domain of our investigation.  
 This lack could also be explained by a wide  distribution of NPBLs  cutoff energies. \\
 The genuine absence of a cutoff would suppose that different emission processes arise within the NPBLs class.\\
 Nevertheless, the cutoff argument cannot  explain the absence of polarized broad lines in the NPBLs sources. 
 Another way to explain the existence of two Sy 2 types is by the study of the central
  engine activity,  explored in the next subsection.
\subsection{The central engine activity}
  We 
show in this study 
 that  the NPBLs countrate at 15 keV is two
  times weaker than PBLs (section 4),  the NPBLs spectrum is softer
   than Sy 1 and PBLs (see 4.3.1 and 4.3.2) and in particular the NPBLs of our sample are less luminous (see 2.3.2). 
   Therefore, the central engine  of NPBLs appears weaker than PBLs. \\
   We speculate that Sy 2 galaxies without polarized broad lines correspond to a state in which the accretion flow is weak and devoid
  of  line emitting clouds. This flow may be insufficient  to produce a large UV flux which in turn would
    explain why Sy 2 without PBL cannot cool through 
   Compton processes between the electron population and the UV photons, thus explaining 
   the absence of cutoff in the high energy emission.   
 \subsection{The absorption/environment role in Seyfert 2 classification}
The absence of  polarized broad lines and of a cutoff in NPBLs galaxies cannot both be  explained by 
 very high
 column densities. Indeed, we consider only Compton-\textit{thin} Sy 2 galaxies  thus sources slightly influenced
  by absorption. 
  The  argument  according  to  which  the  column density of our NPBLs would be
  underestimated is improbable here, since
   this would imply a 
  large divergence between the average emission of Sy 1 and Sy 2 without PBL in the 15-40 keV domain, that is not observed.\\
The absence  of polarized broad lines in NPBLs may be due to either the lack of a scattering medium above the absorbing gas 
 or to the genuine absence of line emitting clouds in the nuclear regions. \\ 
  However, the absence of a scattering medium  alone cannot 
  explain all the differences seen in the continuum emission of the two Sy 2 types. 
\subsection{Implications}
We summarize in Table 5 the characteristics of   Seyfert 2 galaxies with and without polarized broad lines.
\begin{table*}[!hbt]
 \begin{center}
   \caption{Emission properties of  PBLs and NPBLs of our sample}
\begin{tabular}[!hbt]{l | l } 
\hline
\hline
Sy 2 Class & Characteristics  \\
\hline
PBLs & Cutoff at $\sim$ 80 keV \\ 
 & lower spectral index than NPBLs \\
 & Spectral similarities with Sy 1 galaxies  \\
\hline
NPBLs & no constrained cutoff  \\
 & softness of the emission\\
  & central engine activity weaker than PBLs  \\
  & not more absorbed than PBLs   \\
\hline
\end{tabular}
\end{center}
\end{table*}
The PBLs and  Sy 1 galaxies of our sample have similar physical conditions in the inner regions of their sources.
 Consequently,  the  observed  differences  in  the line properties and in X-ray
 emission can be due to a greater 
  absorption in PBLs compared to  Sy
  1 galaxies, which agrees with unified models.\\
The NPBLs, however,  are not only hidden Sy 1 nucleus since the absorption argument is unable to
 reproduce the different spectral properties found in this study.  
 To confirm this assertion and to  allow  a meaningful analysis of the
    central engine signatures of the different types of Sy galaxies, we  need a larger sample of Sy 2 
    having both spectropolarimetric
   observations and high quality X-ray data. \\
We conclude that while Seyfert 2 galaxies with polarized broad lines have  continuum emission properties that
 match those of Seyfert 1 galaxies,
Seyfert 2 galaxies without  PBL do seem to present genuine differences in their nuclear properties.

\begin{acknowledgements}
The author wishes to sincerely thank the anonymous referee for the fruitful     
report he provided, M. Gaber for his advice and P. Veron for the useful         
comments                                                                        
on the "zoology" of  Seyfert  galaxies.     
\end{acknowledgements}

\bibliographystyle{apj}
\bibliography{biblio}

\appendix
\section{Average spectra fitted with the different models}
We present the average count spectrum of each class fitted by the models applied in this study.
\begin{figure}[!h]
\centering
\includegraphics[width=6cm, angle=-90]{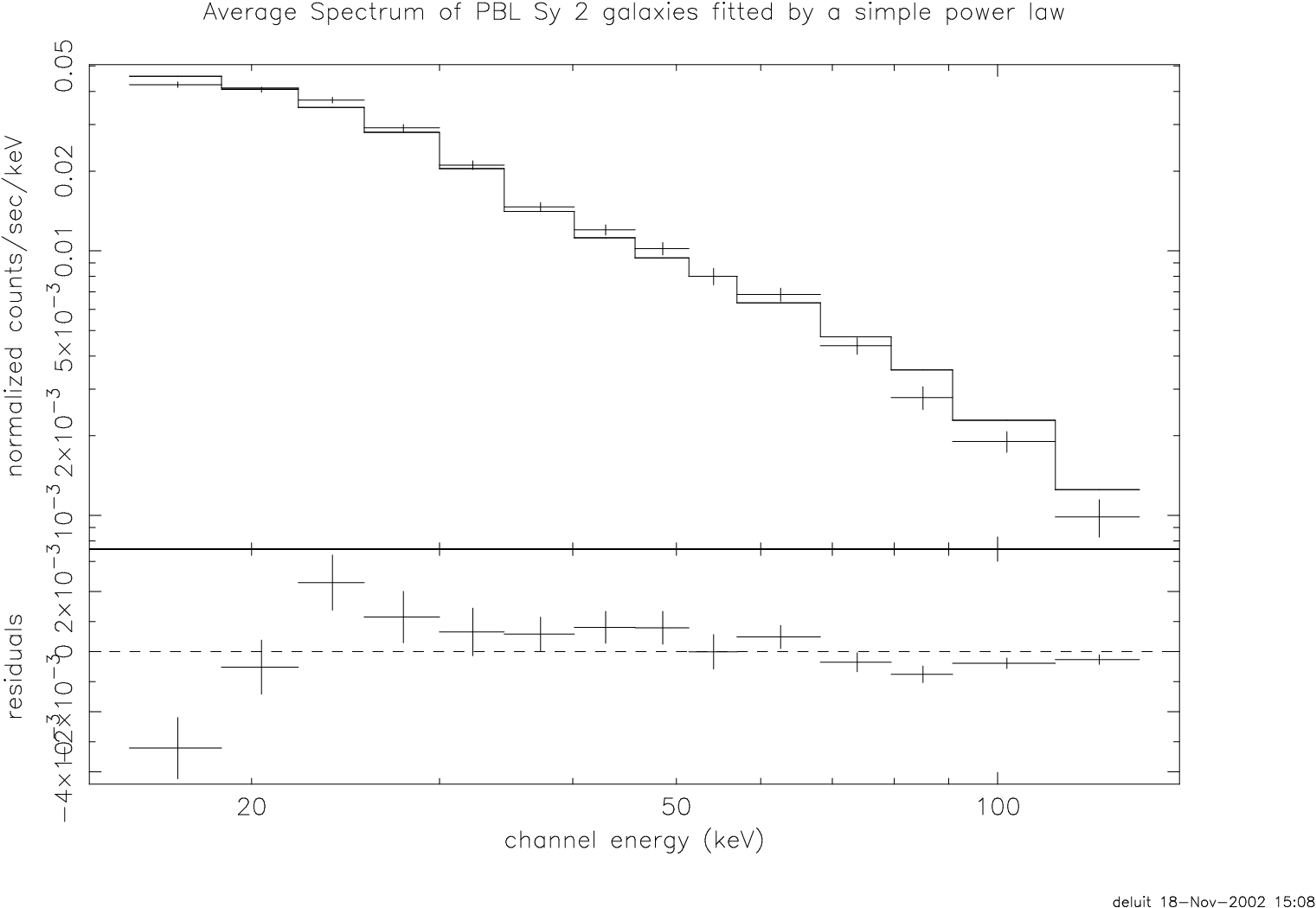}
\caption{Average spectrum of Seyfert 2 galaxies with polarized broad lines fitted by a  power law model and residuals}
\end{figure}

\begin{figure}[!h]
\centering
\includegraphics[width=6cm, angle=-90]{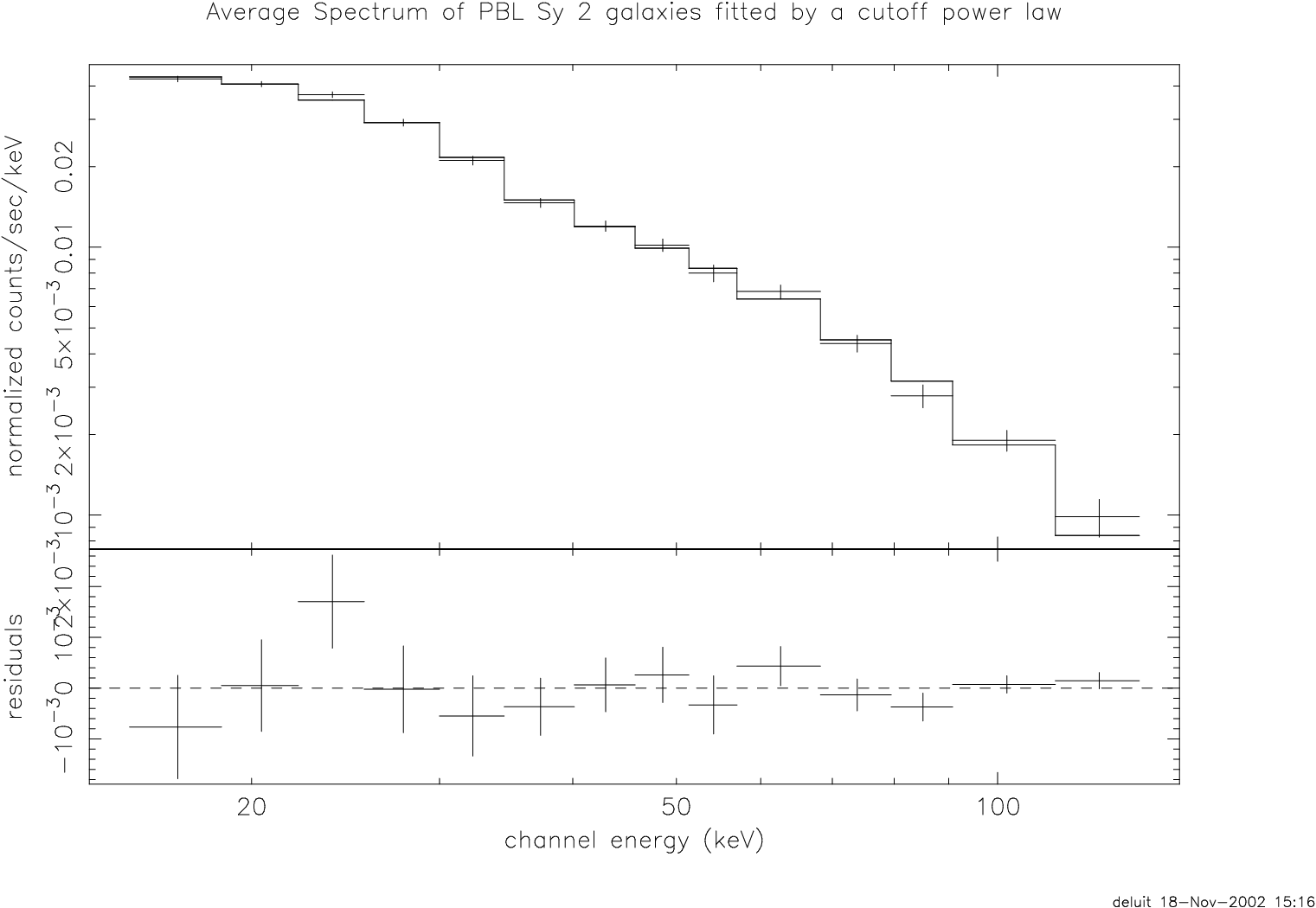}
\caption{Average spectrum of Seyfert 2 galaxies with polarized broad lines fitted by a cutoff power law  model and residuals}
\end{figure}
\begin{figure}[!h]
\centering
\includegraphics[width=6cm, angle=-90]{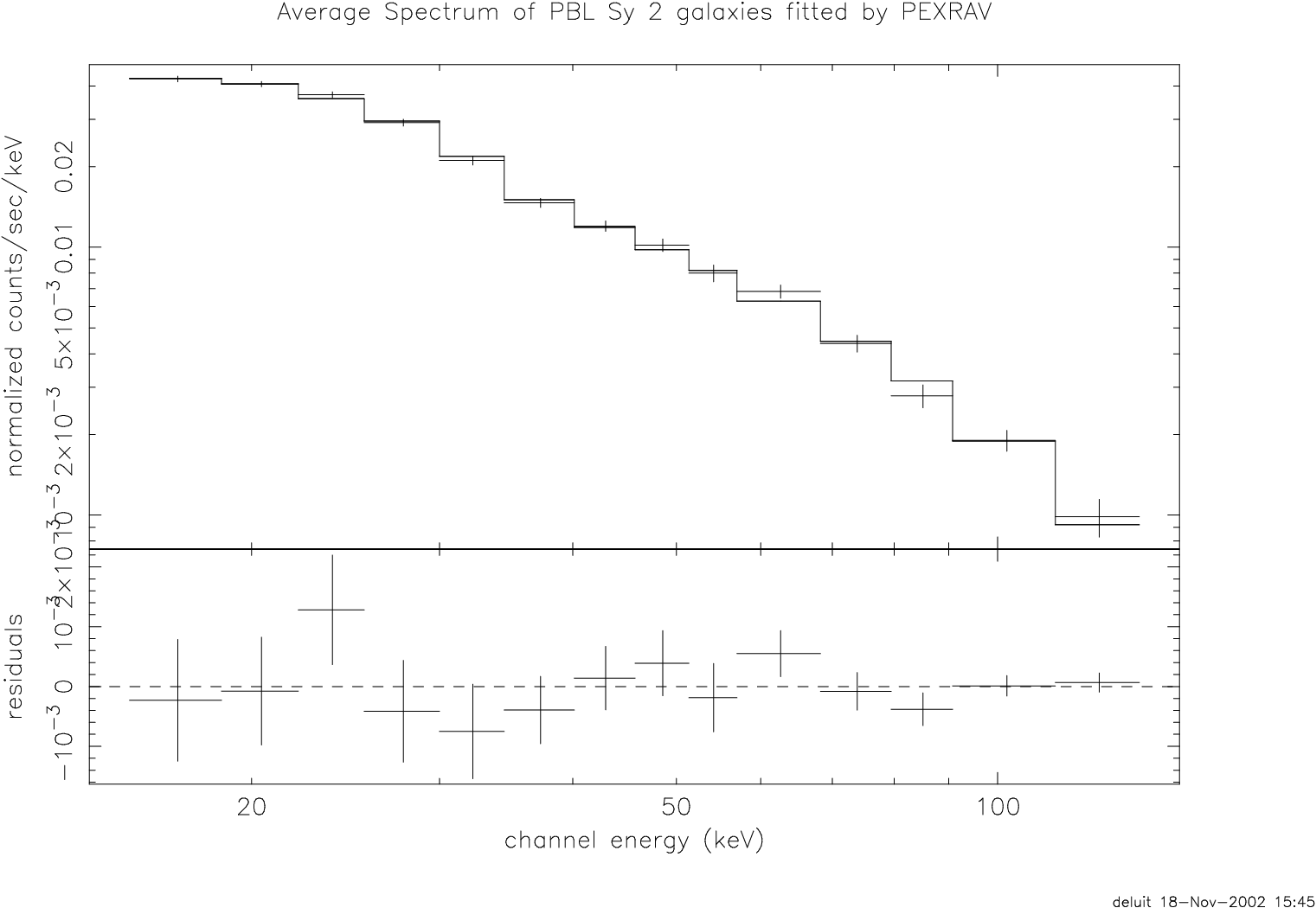}
\caption{Average spectrum of Seyfert 2 galaxies with polarized broad lines fitted by the PEXRAV model (cos $\theta$=0.45) and residuals}
\end{figure}

\begin{figure}[!h]
\centering
\includegraphics[width=6cm, angle=-90]{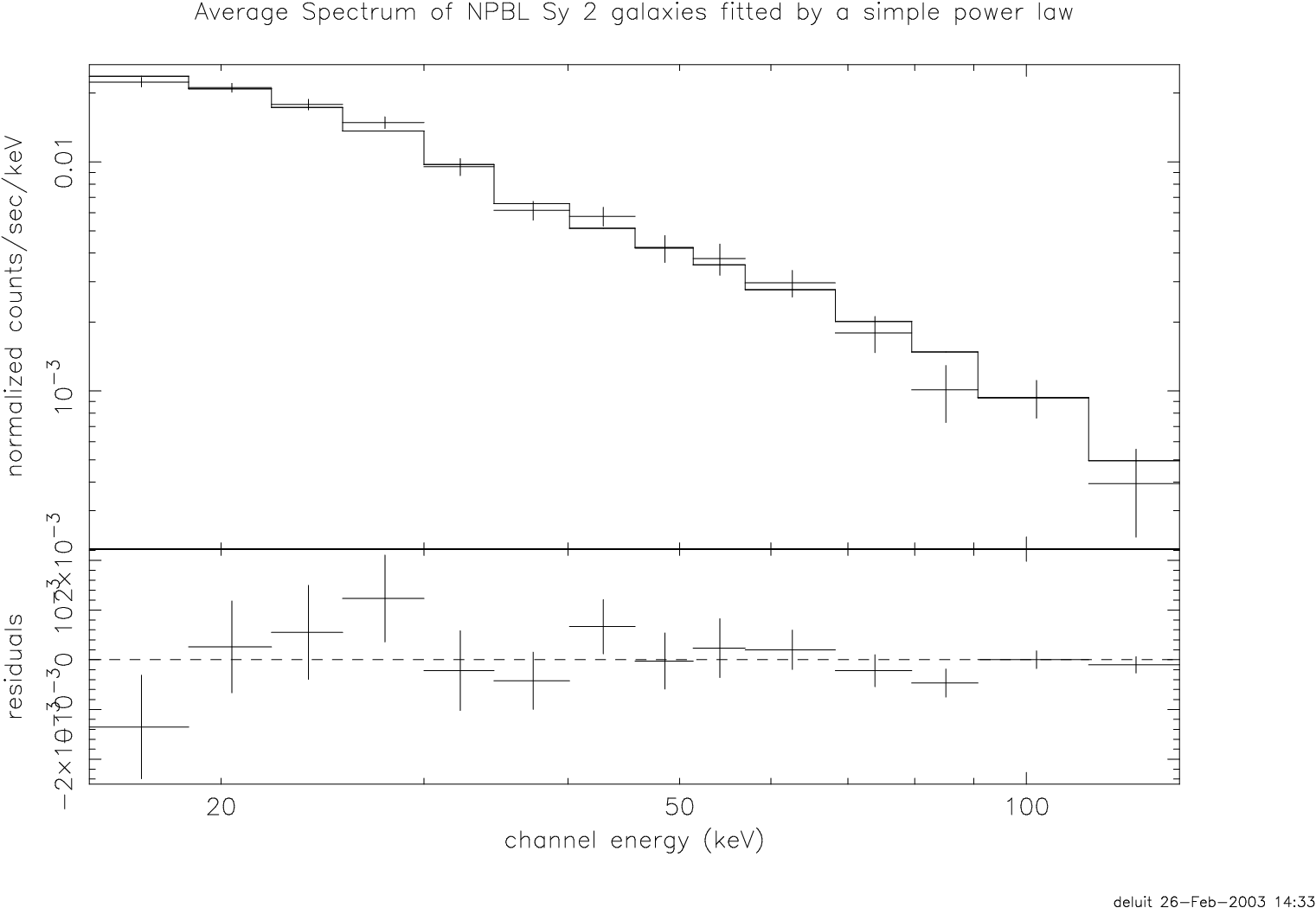}
\caption{Average spectrum of Seyfert 2 galaxies without  polarized broad lines fitted by a  power law  model and residuals}
\end{figure}

\begin{figure}[!h]
\centering
\includegraphics[width=6cm, angle=-90]{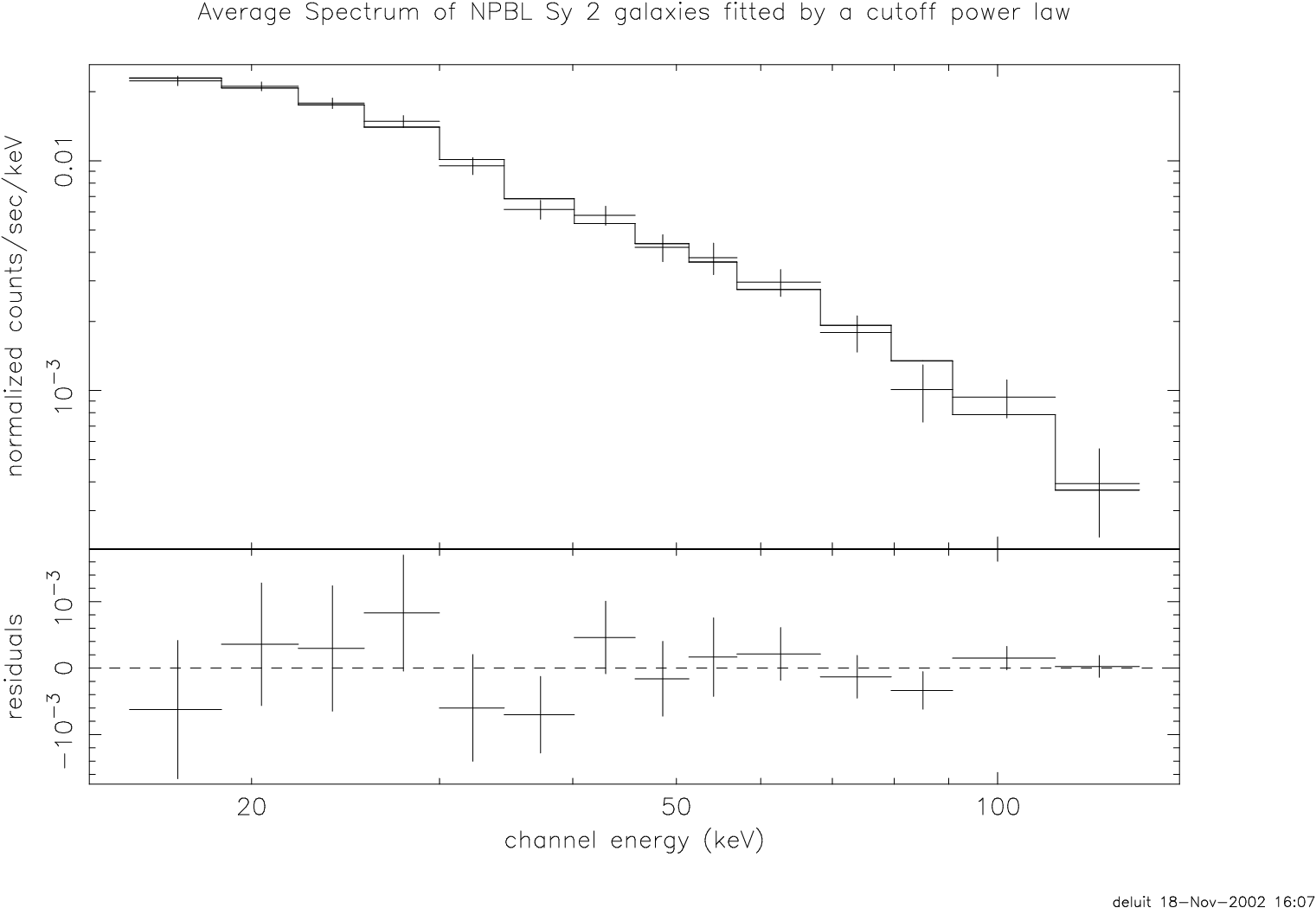}
\caption{Average spectrum of Seyfert 2 galaxies without polarized broad lines fitted by a cutoff power law model  and residuals}
\end{figure}
\begin{figure}[!h]
\centering
\includegraphics[width=6cm, angle=-90]{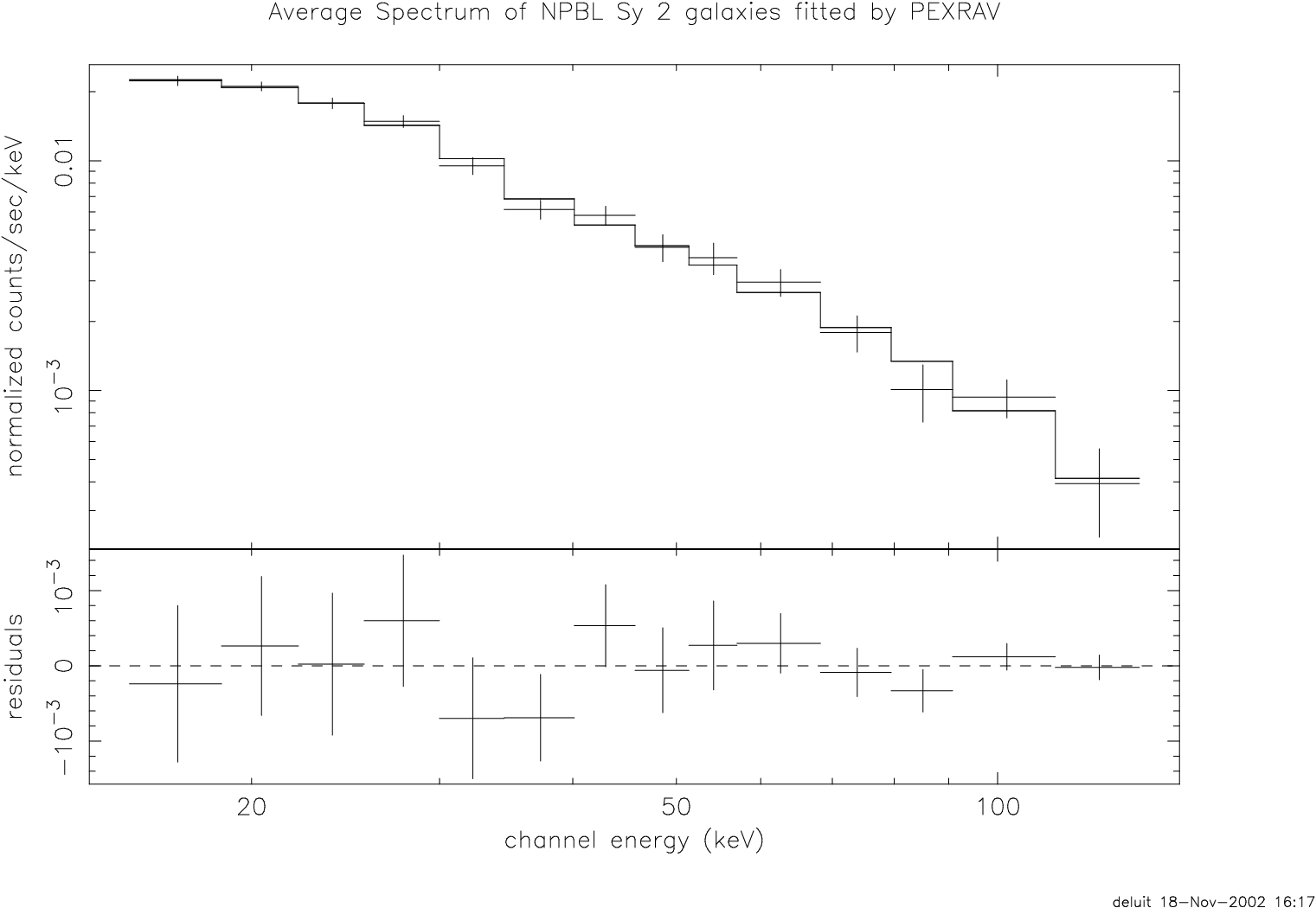}
\caption{Average spectrum of Seyfert 2 galaxies without polarized broad lines fitted by the PEXRAV model (cos $\theta$=0.45) and residuals}
\end{figure}
\end{document}